\documentclass[%
reprint,
amsmath,amssymb,
aps,
prl,
]{revtex4-2}

\usepackage{graphicx}
\usepackage{dcolumn}
\usepackage{bm}
\usepackage{physics}
\usepackage{amsmath}
\usepackage{xcolor}
\usepackage[utf8]{inputenc}
\usepackage[T1]{fontenc}
\usepackage{textcomp}
\usepackage{hyperref}
\usepackage{mathtools}

\renewcommand{\i}{\mathrm{i}}

\renewcommand{\b}{\mathrm{b}}

\renewcommand{\t}{\mathrm{t}}

\newcommand{\p}{\mathrm{p}}
\newcommand{\f}{\mathrm{f}}
\newcommand{\s}{\mathrm{s}}
\newcommand{\g}{\mathrm{g}}

\newcommand{\pati}[1]{}% % \refstepcounter{paticounter}% \noindent\textcolor{blue}{\textbf{\thepaticounter.\ #1}}

\makeatletter
\newcommand{\prlappendix}[1]{%
  \refstepcounter{section}%
  \setcounter{equation}{0}%
  \par\noindent\textit{Appendix \thesection: #1---}%
}
\makeatother

\newcommand{\vg}{v_{\mathrm g}}
\newcommand{\zi}{z_{\mathrm{int}}}
\newcommand{\ts}{t_{\mathrm s}}
\newcommand{\tb}{t_{\mathrm b}}
\newcommand{\te}{t_{\mathrm e}}
\newcommand{\tfoc}{t_{\mathrm f}}
\newcommand{\zfoc}{z_{\mathrm f}}

\begin{document}

\title{Space-Time Lensing by Accelerated Interfaces in Dispersive Media}

\author{C\'edric D'Exelle}
\author{Klaas De Kinder}
\author{Christophe Caloz}
\email{christophe.caloz@kuleuven.be}
\affiliation{WaveCore, KU Leuven, Leuven, Belgium}

\begin{abstract}
We present space-time lensing by synthetic accelerated interfaces in dispersive media. Frequency-dependent group velocities map spectral components of a pulse onto distinct trajectories in the longitudinal space-time plane, while an accelerated interface reshapes their frequencies through time-dependent Doppler shifts. We develop an inverse-design procedure for the interface trajectory that focuses a quasi-monochromatic wave packet at a prescribed space-time event for an arbitrary dispersion relation, then extend the construction to polychromatic pulses. In its time-reversed form, the operation collimates a broadband pulse into a narrow-band wave. Moreover, cascading the forward and reverse operations enables a broadband pulse to propagate through an otherwise dispersive channel in a spectrally compressed state before being refocused. Full-wave simulations validate the focusing construction and demonstrate dispersion suppression.
\end{abstract}

\maketitle
\pati{Lensing Background}
Lensing is the controlled bending and redistribution of waves by inhomogeneous media or artificial structures, enabling focusing, collimation, magnification and imaging. The vast majority of lenses to date have been \emph{spatial lenses}---spatially varying optical-delay devices realized with materials of various shapes and compositions~\cite{BornWolf,SalehTeich}. Although spatial lenses dominate lensing technology, \emph{temporal lenses} have also been reported as their dual counterparts~\cite{Kolner_temporal_imaging,Kolner_1994_spacetime_duality_and_temporal_imaging}---temporally varying optical-delay devices realized by modulating a host medium or waveguide in time. Unlike spatial lenses, which rely on diffraction and quadratic spatial phasing, temporal lenses combine quadratic temporal phase modulation with dispersive propagation, thereby specifically enabling pulse compression, temporal imaging and ultrafast waveform manipulation~\cite{Bennett_principles1,Bennett_principles2,Goda_2009,Li_temp_imaging,Zhu_2022_spectral_control}.

\pati{Gap}
Recent advances in space-time-modulated metamaterials---engineered media whose electromagnetic properties are varied in both space and time---have established the potential use of moving~\cite{Lampe_EM_waves_moving_ionization_front,Reed_reversed_Doppler_effect,Biancalana_2007_propagation_in_spatiotemporal_structures,Gaafar_front_induced_transitions,Caloz_metamaterials1,Caloz_metamaterials2,Caloz_GSTEM} and, specifically, accelerating~\cite{Bahrami_2023_accelerated_modulation,DeKinder_acceleration,Bahrami_pulse_shaping} refractive-index boundaries as wave-control elements. This opens a route to lensing that differs from both spatial index shaping and temporal phase modulation. When a wave crosses an accelerating interface, its frequency is Doppler-shifted by an amount that depends on the interface velocity at the time of crossing. Although moving space-time interfaces have been studied extensively as frequency-converting scatterers~\cite{Sloan_two_photon,DeckLeger_2024_spacetime_transitions,Bahrami_spacetime_wedges,DeKinder_DoPA,Bahrami_pulse_shaping,DeKinder_acceleration,DeKinder_ST_dispersive}, their use as geometrically designed lensing elements for dispersive pulses remains largely unexplored. The resulting question is whether the motion of an interface can be tailored to produce prescribed pulse transformations, such as temporal focusing, spectral compression and suppression of dispersive spreading.

\pati{Contribution}
Here, we address this question by developing a trajectory-based lensing mechanism. In a dispersive medium, frequency determines the group velocity and hence the slope of a wave component in the longitudinal spatial plane. A moving interface provides control over that frequency through space-time-dependent Doppler shifting. By tailoring the interface trajectory, one can therefore control the evolution of spectral components after crossing the boundary. We show that such a scheme yields a space-time lens capable of temporal focusing in media with arbitrary dispersion relations, conversion of broadband pulses into narrow-band waves, and suppression of dispersive pulse spreading. In this lens, the functionality is encoded in the kinematics of the moving boundary rather than in a static index profile or a prescribed temporal phase modulation.

\begin{figure}[h!]
    \centering
    \includegraphics[width=0.98\linewidth]{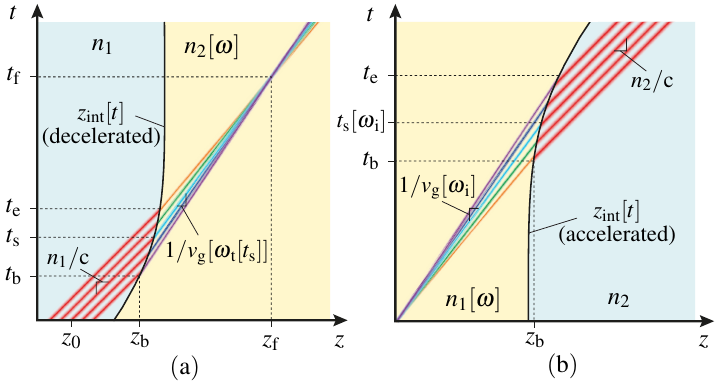}
    \caption{Concept of space-time lensing, with different frequencies corresponding to different trajectories and encoded by different colors. (a)~Focusing. (b)~Collimation.}
    \label{Fig:concept}
\end{figure}

\pati{Focusing Concept}
We first consider space-time \emph{focusing}, illustrated in Fig.~\ref{Fig:concept}(a). An initially quasi-monochromatic wave packet crosses an accelerated interface, which converts it into a polychromatic wave packet through a space-time-dependent Doppler shift. The resulting spectral components then propagate through a dispersive medium with different group velocities and therefore follow different trajectories in the $(z,t)$ plane. By appropriately designing the interface acceleration, these trajectories are arranged to converge at a prescribed focal point $(\zfoc,\tfoc)$, thereby concentrating the pulse in both space and time. This operation corresponds to the original space-time lens suggested by Ostrovskii in 1975~\cite{Ostrovskii_1975} without proof or analysis.

\pati{Collimation Concept}
The inverse operation is space-time collimation, shown in Fig.~\ref{Fig:concept}(b). Here, an initially polychromatic wave packet propagates through a dispersive medium, causing its spectral components, having different group velocities, to separate and follow different trajectories. Upon crossing an accelerating interface, the components undergo different Doppler shifts determined by their respective scattering events. The interface trajectory is chosen so that all components emerge with the same frequency after the space-time-dependent Doppler shift. Space-time collimation thus eliminates the pulse spreading induced by dispersion and compresses the wave spectrum, analogous to how a conventional collimating lens transforms diverging rays into a parallel beam.

\pati{Dispersion Suppression}
These two operations constitute the canonical building blocks of space-time lensing, analogous to focusing and collimation in conventional optics. More generally, they reveal how accelerated interfaces can be used to engineer spectral trajectories in dispersive media. As an example, we show below that cascading a collimating space-time lens with a focusing space-time lens enables dispersion suppression: a pulse is first collimated [Fig.~\ref{Fig:concept}(b)], then propagated without further spreading, and subsequently refocused [Fig.~\ref{Fig:concept}(b)]. This allows the effects of dispersion to be circumvented through boundary kinematics rather than through conventional dispersion-management techniques.

\pati{Detailed Setup and Concept}
We now formulate the corresponding trajectory-design problem, with the help of Fig.~\ref{Fig:concept}. We consider a one-dimensional pulse propagating along $z$, with linearly $x$-polarized electric field~$E_x$. In the geometric-optics regime, where the carrier period $T_\i$ is much shorter than the pulse duration, the pulse can be decomposed into infinitesimal wave groups, or rays, each propagating with the frequency-dependent group velocity $\vg[\omega]$. Consequently, dispersion associates frequency with a trajectory slope in the $(z,t)$ plane. Each ray reaches the accelerated interface at a unique scattering event $(\zi[\ts],\ts)$, where $\ts$ is the time at which this event occurs. This interface need not be a moving material boundary; in practice it may be implemented as a longitudinal traveling-wave modulation of the refractive index~\cite{Caloz_GSTEM,Bahrami_2023_accelerated_modulation}. At each scattering event, the ray undergoes a Doppler frequency transition, $\omega_\i(\ts)\longrightarrow\omega_\t(\ts)$, where $\omega_\t$ is the frequency of the transmitted ray, depending on the velocity of the interface. The general design problem is therefore the following: given an incident pulse, a dispersion relation, and a desired transmitted wave evolution, determine the interface trajectory $\zi(t)$. Focusing then corresponds to a constant incident frequency and transmitted rays that meet at the focal point $(\zfoc,\tfoc)$, whereas collimation corresponds to a varying incident frequency and a constant transmitted frequency.

\pati{Trajectory Calculation}
We start by solving the space-time focusing problem and subsequently use the same construction for collimation. For simplicity, consider a quasi-monochromatic incident pulse of frequency $\omega_\i$ in a nondispersive region with refractive index~$n_1$. Before scattering, its rays are parallel and satisfy $z[t]=z_0+(c/n_1)t$, where $z_0$ is the ray position at $t=0$, as seen in Fig.~\ref{Fig:concept}(a). After a ``ray'' crosses the interface at time $\ts$, it propagates in region~2 according to
$z[t;\ts]=\zi[\ts]+\vg\!\left[\omega_\t[\ts]\right](t-\ts).$
Space-time focusing requires every transmitted ray to pass through the prescribed focal event:
\begin{equation}
\label{Eq:focusing_condition}
    \zi[\ts]+\vg\!\left[\omega_\t[\ts]\right](\tfoc-\ts)=\zfoc,\; \forall\,\ts.
\end{equation}
Differentiating Eq.~\eqref{Eq:focusing_condition} with respect to $\ts$ and using the chain rule yields then the focusing equation
\begin{equation}
\label{Eq:focusing_equation}
    c\beta[\ts]+\frac{d\vg[\omega_\t]}{d\omega_\t}\frac{d\omega_\t[\ts]}{d\ts}(\tfoc-\ts)-\vg[\omega_\t[\ts]]=0,
\end{equation}
where $\beta[t]=d\zi/d(ct)$ is the normalized interface velocity. Equation~\eqref{Eq:focusing_equation} contains two unknown functions, $\beta[\ts]$ and $\omega_\t[\ts]$, which are related by the moving-interface phase-matching condition~\cite{DeckLeger_2024_spacetime_transitions,DeKinder_ST_dispersive}; see Appendix~\ref{App:frequency_transitions},
\begin{equation}
\label{Eq:Doppler}
    \left(\beta[t]n_1[\omega_\i]-1\right)\omega_\i = \left(\beta[t]n_2[\omega_\t[t]]-1\right)\omega_\t[t].
\end{equation}
Solving Eq.~\eqref{Eq:Doppler} for $\beta[t]$ and substituting the result into Eq.~\eqref{Eq:focusing_equation} produces a first-order differential equation for~$\omega_\t[\ts]$. Once a dispersion relation $n_2[\omega]$ and an initial transmitted frequency $\omega_\t[t_\b]$ are specified, this equation can be solved numerically. Substituting this solution back into equation~\eqref{Eq:Doppler} then determines $\beta[\ts]$, and integration gives the sought after interface trajectory
\begin{equation}
    \zi[\ts]=z_b+c\int_{\tb}^{\ts}\beta[t]\,dt.
\end{equation}
The free initial position $z_b$ and beginning of the acceleration $\tb$ have to be chosen so that the incident pulse encounters the interface during the prescribed acceleration duration $[\tb,\te]$, which acts as the temporal aperture. The obtained transmitted frequencies must moreover be causal, namely $\vg[\omega_t]>c\beta$~\cite{DeckLeger_2024_spacetime_transitions}. Thus, apart from the existence of a physical transmitted branch, the construction is not restricted to quadratic group-velocity dispersion.

\pati{Application to a Lorentz Medium}
As a representative example, we consider the lossless Lorentz dispersion relation $n_2[\omega]=\sqrt{1+\omega_\p^2/(\omega_0^2-\omega^2)}$, which yields, upon substitution in Eq.~\eqref{Eq:Doppler}, an explicit group-velocity function and hence an explicit form of Eq.~\eqref{Eq:focusing_condition}, derived in Appendix~\ref{App:explicit_focusing_eq}. Figure~\ref{fig:validation} shows the interface trajectory obtained by solving this focusing equation numerically. Figure~\ref{fig:validation}(a) displays the transmitted frequencies and corresponding ray trajectories, while Fig.~\ref{fig:validation}(b) shows a full-wave simulation of the same lens compared with the analytical propagation limits from Fig.~\ref{fig:validation}(a).

\begin{figure}[h!]
    \centering
    \includegraphics[width=\linewidth]{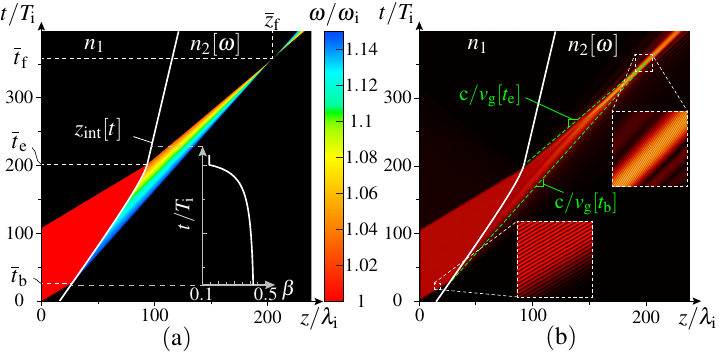}
    \caption{Computed space-time focusing [Fig.~\ref{Fig:concept}(a)] of a quasi-monochromatic incident pulse with frequency $\omega_\i$ by the synthesized accelerated interface $z_\mathrm{int}[t]$, separating a nondispersive medium with refractive index $n_1=1$ from a Lorentz-dispersive medium with refractive index $n_2[\omega]$. The $\omega_\i$-normalized frequencies
are $\overline{\omega}_\p=0.5$, $\omega_0=1.4$, $\overline{\omega}_\t[t_\mathrm{b}]=1.15$ and the $T_\i$-normalized times ($T_\i = 2\pi/\omega_\i=\lambda_\i/c$) $\overline{t}_{\mathrm{b}}=26$, $\overline{t}_{\mathrm{e}}=201$ and $\overline{t}_\f=356$. The incident pulse has a block shaped envelope with exponential edges and duration of $100 T_\i$, centered around $t=55 T_\i$ at $z=0$.
(a)~Theoretical color-coded trajectories of the incident (unique) and scattered (multiple) frequencies. The normalized interface velocity obtained from solving Eq.~\ref{Eq:focusing_equation} is shown in the inset. 
(b)~Full-wave---FDTD---simulated electrical field magnitude compared with the analytically predicted propagation limits, represented by green dashed lines.}
    \label{fig:validation}
\end{figure}

\pati{Discussion}
The interface is modeled as an abrupt refractive-index transition, an appropriate approximation when its rise time is short compared to the carrier period~\cite{Lustig_2023,Li_ST_GRIN}. Since the conventional FDTD update used to obtain Fig.~\ref{fig:validation}(b) does not reproduce moving-boundary scattering amplitudes exactly~\cite{Bahrami_2023_generalized_FDTD_scheme_for_arbitrary_Spacetime_configurations, Xu_2025_FDTD}, this figure only validates the frequency transitions and wave redistribution but does not correspond to the correct physical scattered wave amplitudes. Reflection is kept negligible by approximate impedance matching over the transmitted frequency band, which also reduces the error on the scattered wave amplitudes.

\pati{Pulse Regime Extension}
The assumption of a quasi-monochromatic incident wave is made only to obtain a unique transmitted frequency as a function of scattering time. Space-time focusing can be extended to broadband incident pulses by allowing the initial medium~$n_1[\omega]$ to be dispersive as well. After sufficient predispersion, each incident frequency reaches the interface at a distinct scattering time, so that $\omega_\i[\ts]$ again becomes single valued. The focusing condition~\eqref{Eq:focusing_condition} and the transition law~\eqref{Eq:Doppler} remain valid, but the unknown functions $\omega_\i[\ts]$ and $\omega_\t[\ts]$ must now be found from two coupled differential equations given in Appendix~\ref{App:polychromatic}. Efficient focusing requires opposite signs of $d\vg/d\omega$ in the two media so that the component that is fastest before the interface becomes slowest after it, and conversely.

\pati{Collimation}
This same dispersive preparation also provides the starting point for space-time collimation, the time-reversed operation of space-time focusing. Once the broadband incident pulse has been separated so that each frequency reaches the accelerating interface at a different scattering time, the interface trajectory can instead be designed to impose $\omega_\t[\ts]=\omega_\f$ for every incident component. The process then converts a short broadband pulse into a long narrowband wave. Given the incident band $[\omega_{\min},\omega_{\max}]$, the dispersion relation of the incident medium, and the desired output frequency~$\omega_\f$, the inverse problem yields a differential equation for $\omega_\i[\ts]$, derived in Appendix~\ref{App:collimation}. Substituting its solution into Eq.~\eqref{Eq:Doppler} gives the required interface velocity $\beta[\ts]$, which is again integrated to obtain the interface trajectory.

\pati{Dispersion Suppression}
The practical value of collimation becomes apparent when it is combined with a subsequent focusing operation. A broadband pulse can first be collimated to a narrowband wave~[Fig.~\ref{Fig:concept}(b)], then transmitted through an otherwise dispersive channel, and finally refocused to restore the original short pulse~[Fig.~\ref{Fig:concept}(a)]. The spectral compression from the collimation suppresses pulse spreading both by reducing the bandwidth sampled by the channel and, where useful, by shifting the narrow band away from a material resonance. The focusing lens must be the space-time-reversed counterpart of the collimating lens and must be synchronized so that the collimated wave passes through it during the acceleration period. Because the channel is traversed in a narrowband state, this procedure circumvents the group delay from the dispersive channel. Unlike conventional dispersion-compensation schemes, it therefore does not rely on dispersion-engineered elements, such as compensating fibers or chirped Bragg gratings, with a designed opposite group-delay profile~\cite{Agrawal_fiber_optics,Hill_1994_Bragg_grating}. It also avoids the delicate balance between optical nonlinearity and dispersion required for soliton-based transport~\cite{Agrawal_nonlinear_fiber_optics}.

\begin{figure*}[t]
    \centering
    \includegraphics[width=0.9\textwidth]{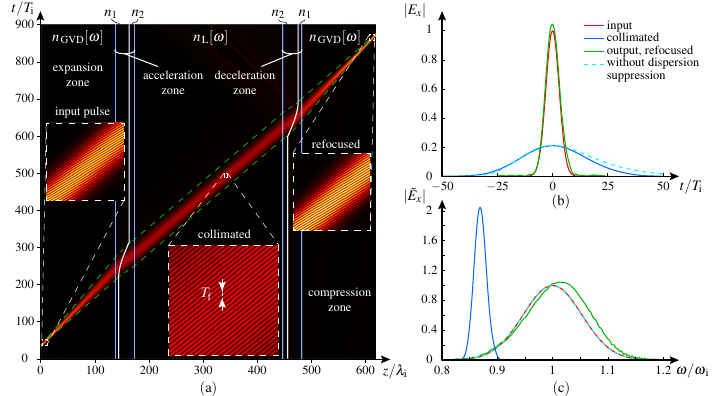}
    \caption{Dispersion suppression of a broadband pulse in a Lorentz-dispersive medium by first space-time collimating the pulse~(Fig.~\ref{Fig:concept}(b)) before propagation in the dispersive medium ($n_\mathrm{L}$) and then refocusing the collimated wave~(Fig.~\ref{Fig:concept}(a))\footnote{Here, the collimation and focusing operations are done in two steps for easy combination of analytical (two outer regions, with $n=n_\mathrm{GVD}$) and numerical (central regions, with $n= n_1,n_2,n_\mathrm{L}$) results.}. The $\omega_\i$-normalized Lorentz dispersion parameters
    are $\overline{\omega}_\p=0.7$, $\overline{\omega}_0=1.6$, the GVD parameters in~\eqref{Eq:quadratic_disp} are $\beta_1 = 1.5$, $\beta_2 = 0.15$ and the refractive indices of the acceleration regions are $n_1=1.5$ and $n_2=1$. The GVD elements are $137\,\lambda_\i$ wide while the Lorentz medium is $275\,\lambda_\i$ wide. The input gaussian spectrum lies between $\overline{\omega}_\mathrm{min}=0.87$ and $\overline{\omega}_\mathrm{max}=1.13$ with the amplitude smaller than $0.05$ outside this region. The pulse is centered around $t=37\, T_\i$ at $z=0$. and the $T_\i$-normalized start and end times are $\overline{t}_{\mathrm{b}}=220$ and $\overline{t}_{\mathrm{e}}=310$. 
    (a)~Field magnitude $|\Re{E_x}|$ obtained from FDTD simulation compared with analytically predicted propagation paths, represented by green dashed lines. (b)~Envelopes $|E_x|$ (with center set to $t=0$) and (c)~spectra of the input wave at $z=0$, the collimated wave at $z=300\lambda_0$ and the refocused wave at $z=z_\f$; for comparison, the wave envelope after propagation through the dispersive medium without dispersion suppression is also shown.}
    \label{Fig:dispersion suppression}
\end{figure*}

\pati{Pulse Recovery Requirements}
Exact recovery of the input waveform requires not only impedance-matched interfaces to avoid loss, but also a quadratic (GVD) dispersion relation in the dispersive sections of the space-time lenses that perform the temporal expansion and recompression, specifically,
\begin{equation}\label{Eq:quadratic_disp}
k[\omega]=k_0+\beta_1(\omega-\omega_0)+\beta_2(\omega-\omega_0)^2.
\end{equation}
This recovery condition lies beyond the ray-optics construction, which fixes the spectral trajectories but does not determine whether the corresponding field amplitudes are restored. During collimation, dispersion maps each frequency to a trajectory slope through the group delay $(dk/d\omega=1/v_\mathrm{g})$; for a nonquadratic dispersion relation, this mapping is nonlinear, so equal spectral intervals are stretched into unequal temporal intervals. The resulting narrowband wave can therefore contain a concentrated main feature accompanied by a long, weak tail. Although the subsequent focusing lens reverses the ordering of the trajectories---mapping the slowest input components to the fastest output components and conversely---it does not, in general, reverse this nonuniform redistribution of amplitude, and the reconstructed pulse is distorted. Quadratic dispersion is special because its group delay is linear in frequency, ensuring uniform spectral expansion and recompression and thereby enabling faithful recovery of the original waveform. This amplitude-level requirement is not apparent from a ray-optics construction alone. Importantly, it applies only to the dispersive sections within the lenses; the intervening transmission channel may remain arbitrarily dispersive.

\pati{Disp. Supp.: Results and Discussion}
Figure~\ref{Fig:dispersion suppression} illustrates the dispersion-suppression scheme for a broadband pulse traversing a Lorentz-dispersive channel. In the simulation, each space-time lens is implemented in a split form: the quadratic-GVD section that performs temporal expansion or recompression is spatially separated from the accelerated interface, which is modeled between two nondispersive media with $(n_1=1.5)$ and $(n_2=1)$. This separation is useful both numerically and experimentally. Numerically, it allows the field evolution in the exact quadratic-GVD sections to be computed analytically, while FDTD is used only for the accelerated interfaces and for propagation through the Lorentz-dispersive channel. Experimentally, it reflects a modular implementation in which the dispersion-engineered element and the nonlinear platform used for accelerated modulation can be optimized independently. Figure~\ref{Fig:dispersion suppression}(a) shows the resulting space-time evolution of the electric field: the broadband input pulse is first expanded and collimated into a narrowband intermediate wave, transmitted through the dispersive channel, and finally recompressed and refocused. Figure~\ref{Fig:dispersion suppression}(b) compares the centered envelopes of the input, collimated, output, and reference pulses, demonstrating recovery of the initial short-pulse envelope after the two lens operations. Figure~\ref{Fig:dispersion suppression}(c) shows the corresponding spectra, confirming the intermediate spectral compression and subsequent restoration of the original bandwidth. In the absence of the two space-time lenses, the same Lorentz-dispersive channel strongly broadens the pulse without altering its spectrum.

\pati{Experimental Prospects}
Experimentally, the space-time lens requires an accelerated longitudinal refractive-index modulation together with controlled dispersion. These ingredients are increasingly available in optically induced moving refractive-index fronts~\cite{Philbin_analog_of_event_horizon,Caspani_2016,Gaafar_reflection_front,Gaafar_front_induced_transitions,Kinsey_2019,Ball_ST_knife_edge,Harwood_ST_diffraction,Zhang_2026_moving_index_front}. Raman-decelerated solitons in optical fibers already provide decelerating index perturbations capable of temporal reflection, focusing, and waveguiding of probe pulses~\cite{Zhang_2022_temporal_reflection}, although with limited control over the acceleration. More versatile acceleration may be obtained from pump-induced index fronts with curved wavefronts: the index perturbation follows the intersection of the pump front with the probe axis. This intersection travels with a velocity that depends on the angle between the pump front and probe axis, and can therefore be varied in time by curving the pump front~\cite{Ball_ST_knife_edge,Caloz_GSTEM,Harwood_2026_synthetic_motion}. These developments suggest that a prototype space-time lens is feasible, with practical limits set by finite rise time~\cite{Zhang_2021_boundary_sharpness}, loss, reflection, modulation depth, and bandwidth, which constrain the temporal aperture and frequency-conversion range but not the trajectory-synthesis principle.

\pati{Conclusion}

In conclusion, we have established accelerated space-time interfaces as a new class of lensing elements for dispersive waves. By combining the frequency-to-trajectory mapping provided by dispersion with the time-dependent Doppler shifts produced by an accelerating boundary, we formulated space-time lensing as an inverse-design problem and derived interface trajectories for prescribed pulse transformations. This framework enables temporal focusing for general dispersion relations, spectral collimation through the time-reversed operation and dispersion-suppressed transport through cascaded lenses. Unlike conventional spatial or temporal lenses, the proposed functionality is encoded in the kinematics of a synthetic boundary rather than in a static index profile or prescribed temporal phase modulation. Beyond the specific demonstrations presented here, this trajectory-based approach provides a general platform for controlling the spectral and temporal evolution of waves and may enable new forms of pulse compression, waveform recovery, dispersive-channel transmission and space-time signal processing in optical, microwave and other wave systems. 

\bibliography{ST_Lens}

\appendix
\setcounter{section}{0}
\renewcommand{\thesection}{\Alph{section}}
\renewcommand{\theequation}{\thesection\arabic{equation}}

\prlappendix{Frequency transition through an arbitrarily moving interface}\label{App:frequency_transitions}
The frequency transition follows from the phase matching condition across the interface~\cite{DeckLeger_2024_spacetime_transitions, DeKinder_ST_dispersive},
\begin{equation}\label{Eq:phase_matching}
    \phi_\i = k_\i z_\mathrm{int}[t] - \omega_\i t = k_\t z_\mathrm{int}[t] - \omega_\t t = \phi_\t.
\end{equation}
Inserting the dispersion relation $k = n_l[\omega]\, \omega/c$, with $l\in\{1,2\}$ for the incident and transmitted wave, respectively, into~\eqref{Eq:phase_matching} yields
\begin{equation}
\label{Eq:phase_subst}
    \left(n_1[\omega_\i]\,\frac{z_\mathrm{int}[t]}{c} - t\right)\omega_\i = \left(n_2[\omega_\t]\,\frac{z_\mathrm{int}[t]}{c} - t\right)\omega_\t.
\end{equation}
Differentiating~\eqref{Eq:phase_subst} with respect to time and defining the normalized interface velocity as $\beta[t]=\frac{dz_{\mathrm{int}}[t]}{d(c t)}$, one obtains the frequency transition law
\begin{equation}
    \left(\beta[t]n_1[\omega_\i]-1\right)\omega_\i=
    \left(\beta[t]n_2[\omega_\t]-1\right)\omega_\t,
\end{equation}
which is Eq.~(2) in the letter. In the design problem, this relation is solved at each scattering time for the dispersion-induced causal transmitted branch satisfying $v_\g[\omega_\t]>c\beta$ \cite{DeckLeger_2024_spacetime_transitions,DeKinder_acceleration}.

\prlappendix{Group velocity and explicit focusing equation for a Lorentz medium}\label{App:explicit_focusing_eq}
In a lossless Lorentz medium, the dispersion relation is $k=n_2[\omega]\,\omega/c$, with
\begin{equation}
    n_2[\omega]=\sqrt{1+\frac{\omega_\p^2}{\omega_0^2-\omega^2}}
\end{equation}
Differentiating with respect to $\omega$ and using the product rule gives then
\begin{equation}
    dk=\frac{\omega\, n_2'[\omega]+n_2[\omega]}{c}d\omega,
\end{equation}
and therefore
\begin{equation}
    v_\g[\omega]\coloneqq\frac{\partial \omega}{\partial k}=
    \frac{c}{\omega\, n_2'[\omega]+n_2[\omega]}
    =\frac{c(\omega_0^2-\omega^2)^2\, n_2[\omega]}{\omega_\p^2\omega_0^2+(\omega_0^2-\omega^2)^2}.
\end{equation}
Substituting this expression and its derivative into the focusing equation, Eq.~\eqref{Eq:focusing_equation}, yields the explicit first-order equation
\begin{equation}
\label{Eq:lorentz_ode}
\begin{aligned}
    & \frac{\omega_\i-\omega_\t}{\omega_\i n_1-\omega_\t n_2[\omega_\t]}
    -\frac{(\omega_\t^2-\omega_0^2)^2\,n_2[\omega_\t]}{(\omega_\t^2-\omega_0^2)^2+\omega_\p^2\omega_0^2} \\
    & + \frac{\omega_\t\omega_\p^2\left(\omega_\t^4+2\omega_\t^2\omega_0^2-3\omega_0^2(\omega_\p^2+\omega_0^2)\right)}{\left((\omega_\t^2-\omega_0^2)^2+\omega_\p^2\omega_0^2\right)^2\, n_2[\omega_\t]}
    (t_\f-\ts)\frac{d\omega_\t}{d\ts}=0.
\end{aligned}
\end{equation}
Equation~(\ref{Eq:lorentz_ode}) can be solved numerically for $\omega_\t$ once $\tb$, $\te$, $t_\f$ and the initial transmitted frequency, $\omega_\t[\tb]$, have been chosen. Substituting this $\omega_\t$ solution into the frequency transition law, Eq.~\eqref{Eq:Doppler}, then leads to the interface velocity $\beta[\ts]$. Finally, integrating this velocity gives the sought after lens trajectory for an initial condition $z_\b = z_\mathrm{int}[\tb]$, which has to be chosen so that the earliest part of the incident pulse reaches the interface at time $\tb$.

\prlappendix{Polychromatic incident pulses}
\label{App:polychromatic}
The monochromatic-input regime in the letter is fairly simple because it involves a unique incident frequency at each scattering time and therefore a unique Doppler $t_\s\mapsto \omega_\t$ mapping. For a genuinely polychromatic incident pulse, that one-to-one mapping breaks as multiple frequencies overlap at each scattering time and thus experience the same interface velocity and Doppler shift. The lens can then no longer prescribe the frequency ordering required for focusing and the intrinsic dispersive spreading of the pulse will overcome the compression from the lens. To maintain the focusing operation, the input medium must also be dispersive so as to provide precompensation. After sufficient dispersive propagation, the incident frequencies are separated in time before they reach the interface so that at each time $\ts$ there is a unique incident and transmitted frequency. The accelerated interface can then be designed to reorder and broaden the transmitted spectrum so that the transmitted rays refocus inside the second medium. In the two-dispersive-medium case, the focusing equation~\eqref{Eq:focusing_condition} needs to be supplemented by a second equation to determine $\omega_\i[\ts]$. This equation arises from considering the generalized scattering time relation $z_0 + v_{\g,1}[\omega_\i]\,\ts=z_{\mathrm{int}}[\ts]$ and differentiating it with respect to $\ts$ to obtain
\begin{equation}\label{Eq:scattering_time_relation}
        \frac{\partial v_{\g, 1}[\omega_\i]}{\partial \omega_\i}\, \frac{\partial \omega_\i}{\partial \ts}\, \ts + v_{\g, 1}[\omega_\i] = c \beta[\ts].
\end{equation}
To find the required interface trajectory, one now has to solve the focusing equation~\eqref{Eq:focusing_equation} and Eq.~\eqref{Eq:scattering_time_relation} simultaneously as a system of coupled differential equations for $\omega_\i[t_\s]$ and $\omega_\t[t_\s]$. Substituting these solutions into the frequency transition relation~\eqref{Eq:Doppler}, gives again the interface velocity, from which the trajectory is found through integration. Focusing of the incident pulse requires the dispersive separation before the interface to be undone after the interface. A useful design rule is therefore that the relevant branches of $v_{\g,1}(\omega)$ and $v_{\g,2}(\omega)$ should have opposite slopes: frequencies that are fast in the incident medium should be mapped to slow transmitted components, and conversely. This reversal allows the initially dispersed spectral components to recombine at the prescribed focal event, with the frequencies altered by the accelerated interface.

\prlappendix{Space-time collimation}
\label{App:collimation}
Space-time collimation is the time-reversed version of the focusing operation. An incident broadband pulse propagates in a dispersive medium so that each incident frequency $\omega_\i$ reaches the accelerated interface at a distinct time $\ts(\omega_\i)$. The interface is then designed so that all of those frequencies are shifted to a single output frequency $\omega_\f$. Given the incident frequency, the required velocity for this shift follows from solving~\eqref{Eq:Doppler} for $\beta[\ts]$ and substituting $\omega_\t$ by $\omega_\f$, which yields
\begin{equation}
\label{Eq:return_velocity}
    \beta[\ts] = \frac{\omega_\f - \omega_\i}{n_2[\omega_\f]\,\omega_\f - n_1[\omega_\i]\,\omega_\i}.
\end{equation}
Then the acceleration that  shifts the entire incident spectrum to $\omega_\f$ is obtained by determining the incident frequency as a function of scattering time $\omega_\i[\ts]$ and substituting it in~\eqref{Eq:return_velocity}. A differential equation for $\omega_\i[\ts]$ was already derived in the previous section, resulting in~\eqref{Eq:scattering_time_relation}. One then simply substitutes~\eqref{Eq:return_velocity} in~\eqref{Eq:scattering_time_relation} and solves this equation numerically to find $\omega_\i[\ts]$. Inserting the solution back in~\eqref{Eq:return_velocity} gives the desired interface velocity, which can then be integrated to obtain the interface trajectory.

\end{document}